\documentclass[twocolumn,showpacs,preprintnumbers,amsmath,amssymb]{revtex4}

\usepackage{graphicx}
\usepackage{dcolumn}
\usepackage{bm}
\usepackage{booktabs}
\usepackage{amsmath}

\newcommand{\fig}[1]{Fig.~\ref{#1}}

\newcommand{\eq}[1]{Eq.~(\ref{#1})}
\newcommand{\ta}[1]{Table~\ref{#1}}
\newcommand{\cp}[1]{Chapter~\ref{#1}}
\renewcommand{\vec}[1]{\mathbf{ #1 }}

\begin{document}


\title{The fraction of Bose-Einstein condensed triplons in $\mathrm{TlCuCl_3}$ from magnetization $M(T,H)$-data.}

\author{Raffaele Dell'Amore}
 \email{dellamore@physik.uzh.ch}

\author{Andreas Schilling}%
 \email{schilling@physik.uzh.ch}
\affiliation{%
 Physik-Institut, University of Zurich, Winterthurerstrasse 190, 8057 Zurich, Switzerland\\
}%
\author{Karl Kr\"amer}%
 \email{karl.kraemer@iac.unibe.ch}
\affiliation{%
Department of Chemistry and Biochemistry, University of Bern, 3000 Bern 9, Switzerland
}%

\date{\today}

\begin{abstract}
$\mathrm{TlCuCl_3}$ is a quantum- spin- $\frac{1}{2}$ system which shows a gap between the singlet ground state of the $\mathrm{Cu^{2+}}$ dimers and the first excited triplet $S^z=+1$ state for magnetic fields $\mu{_0}H~\lesssim~\mu{_0}H_c~\approx~5.5$~T.  At larger magnetic fields the gap is suppressed, and a Bose-Einstein condensation (BEC) of triplets is supposed to occur, leading to a magnetic phase with antiferromagnetic long-range order of the transverse spin components.
\\ In this study we calculate the fraction of condensed magnetic quasiparticles of $\mathrm{TlCuCl_3}$ from magnetization $M(T,H)$-data. At $T=0$ K and in $\mu_0H=$ 6 T, this fraction is $\approx 98\%$ of the total number of triplons. It is independent of the direction of the magnetic field and slightly decreases with increasing magnetic field if we assume the presence of a small intrinsic magnetic background with $S=1$ magnetic moments. 
\end{abstract}

\maketitle

\section{\label{Intro}Introduction\protect\\ }

Low-dimensional quantum-spin systems exhibit a variety of quantum phenomena that have gained much in interest in the last decade \cite{mats,Affl,wats,rueg, Nik}.
\\$\mathrm{TlCuCl_3}$, for example, is a material in which magnetic quasiparticles carrying spin $S~=~1$ (spin triplet states, here called triplons) are believed to form a Bose-Einstein condensate (BEC) above a critical field $\mu{_0}H_c\approx5.5$ T and at low temperatures \cite{rueg,Nik}.  
Meanwhile several other materials have been found that exhibit various features that can be explained within the framework of a condensation of quasiparticles with integer spin \cite{grenier,zapf,jaime, giaandrueg}.
\\ The magnetic properties of $\mathrm{TlCuCl_3}$ are determined by the exchange interactions between the $\mathrm{Cu^{2+}}$ ions  which are arranged in dimer pairs within  $\mathrm{Cu_2Cl_6}$-clusters. 
\\ The magnetic ground state of $\mathrm{TlCuCl_3}$ is a non-magnetic spin singlet that is separated from the first excited triplet state by an excitation gap $\Delta \approx$ 0.7  meV in zero magnetic field. This gap has been measured, for example, by neutron scattering and ESR measurements \cite{cava, oosawa} which revealed that this gap is due to the strong antiferromagnetic interaction $J$= 5.68 meV in the planar dimer of $\mathrm{Cu_2Cl_6}$. The neighboring dimers are coupled by strong interdimer interactions along the double chain and in the (1 0 -2) plane \cite{cava, oosawa2}.
\\  As soon as the external magnetic field $H$ is larger than a critical field $H_c$ with $g\mu_B\mu_0H_c(T=~0)~=~\Delta$ (where $\mu_B$ is the Bohr magneton and $g$ is the Land\'e g-factor), the excitation gap closes due to the Zeeman splitting, and the triplet states $S^z=+1$ are populated, eventually forming the BEC. The 3D interdimer interactions drive this quantum phase transition to finite temperatures leading to a temperature dependent critical field $H_c(T)$. The characteristic off- diagonal long-range order of the BEC
manifests itself in the antiferromagnetic ordering of the spin system in the plane perpendicular to the applied magnetic field \cite{tanaka}.
\\The idea of BEC has already been used quite successfully to explain the transition of "normal" to "superfluid" $^4\mathrm{He}$   \cite{lond,He2}. The strong interactions that exist in liquid $^4\mathrm{He}$ may alter the nature of the transition, however. For instance, while 90-95\% of the particles of an atomic ensemble are in the "superfluid" phase below the transition temperature of an atomic BEC, just a few percent ($\sim$ 9\%) of the Helium-atoms are condensed in superfluid  $^4\mathrm{He}$.
\\ In this paper we focus on the condensed phase of triplons in $\mathrm{TlCuCl_3}$ at magnetic fields $\mu{_0}H_c<~\mu{_0}H<~9$~T and at temperatures down to $T$= 1.9 K. From magnetization $M(T,H)$ measurements we extract the density of condensed triplons at $T$ = 0 K for $\mathbf{H} \parallel b$ and $\mathbf{H} \parallel $ [201]. Taking various possible contributions to the total magnetization into account, we show that the density of triplons forming the condensate is in fact the same for both directions \cite{giaandrueg}.  We also determine the magnetic-field dependence of the fraction of triplons forming the condensate. The quantitative results presented here confirm the scenario of the formation of a \textit{weakly interacting Bose gas} of triplons right above $H_c$ \cite{Nik, sirker}, and we conclude that the interaction increases with increasing particle density, i.e. with increasing magnetic field $H$.

\section{\label{susc}The magnetic susceptibility for $\mathrm{T> 20 K}$}
\begin{figure}
\includegraphics[scale=0.3]{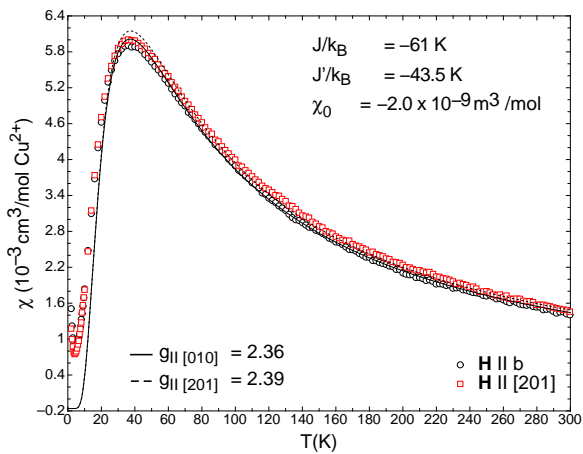}
\caption{\label{suschT} The magnetic susceptibility $\chi(T)$ of $\mathrm{TlCuCl_3}$ with $\mu_0H$ = 1 T applied along the crystallographic $b$-axis and along [201]. The corresponding fits to the data are discussed in the text}
\end{figure}

Magnetic-susceptibility measurements were performed in a commercial PPMS (Physical Property Measurements System, Quantum Design) on a $\mathrm{TlCuCl_3}$ single crystal with mass m = 12.36 mg, for 1.9~K $\leq T  \leq$ 300 K at $\mu_0H$ = 1 T for $\mathbf{H} \parallel b$ and $\mathbf{H} \parallel \mathrm{[201]}$.
\\ The susceptibility $\chi(T)$ of $\mathrm{TlCuCl_3}$ is typical for a low-dimensional spin gap system, showing a well pronounced maximum at $T_{\chi_{max}} \sim$ 36 K and an exponential decrease at low temperatures indicating the existence of a gap $\Delta$ between the ground state and the first excited triplet state, see \fig{suschT}.
For Heisenberg spin systems with identical spin subsystems that are weakly coupled to each other, a good fit to the data in the paramagnetic regime is provided by the molecular mean-field theory (MFT) and its extensions \cite{jonny}. We therefore used this approach within the model of dimers coupled by an effective interdimer coupling $\widetilde{J}$, representing the sum over all exchange coupling constants $J_{kl}$ for a given dimer $k$ interacting with neighboring dimers $l$ \cite{jonny}. An additional temperature-independent diamagnetic term $\chi_{0}$ which contains the orbital diamagnetic core contribution $\chi^{core}$  (including the background contribution of the sample holder) and a paramagnetic Van Vleck contribution $\chi^{VV}$ is also considered.  
\\ Thus we fitted the magnetic susceptibility $\chi(T)$ for temperatures $T>20$ K according to
\begin{subequations}
\begin{equation}
\label{s}
\chi(T)= \chi_{0}+\chi_{MF}(T), 
\end{equation}
with
\begin{equation}
\begin{aligned}
\chi_{0}&=\chi^{core}+\chi^{VV}, 
\end{aligned}
\end{equation}
and
\begin{equation}
\chi_{MF}(T)=\frac{\chi^{dimer}(T)}{1+ \chi^{dimer}(T)\frac{ \widetilde{J}}{N_A g^2 \mu_B^2\mu_0}}.
\end{equation}
Here, 
\begin{equation}
\chi^{dimer}(T) = \frac{N_Ag^2\mu_{B}^2\mu_0}{3 k_B T}\frac{2(S+1)exp({-{\frac{J}{k_BT}})}}{1+ 2(S+1)exp({-{\frac{J}{k_BT}}})}
\end{equation}
\end{subequations}
is the susceptibility of a non-interacting spin-dimer system with single spins $S=\frac{1}{2}$ and the intradimer coupling $J$. $\chi_{MF}$ accounts for the mean-field correction \cite{jonny}.
\\ For a given measured data set of $\chi(T)$-data we therefore used four fitting parameters: $\chi_{0}$, $g$, $J$ and $\widetilde{J}$. We forced the values for $\chi_{0}$, $J$ and $\widetilde{J}$ to be identical for both magnetic-field directions. This restriction is physically reasonable, since these three fitting parameters are independent of the magnetic-field orientation. A small anisotropy of the $g$-factor was considered, however, although it is not predicted by ESR measurements \cite{oosawa}. In fact, the obtained $g$-values for the two investigated crystallographic directions are the same within the error margin, see \ta{para}. The best obtained fits are shown in \fig{suschT}. They yield a good description of the experimental data for $T \geq 20$ K. However, the distinct upturn in $\chi(T)$ at lower temperatures is not at all reproduced by the fits. We believe that this term is intrinsic for $\mathrm{TlCuCl_3}$ \cite{oosawa} and we shall discuss it in more detail in \cp{mag}. Note that the inclusion of a Curie-like term for fitting the data at $T>20$ K does not significantly change the results presented in \ta{para}.

\begin{table}[htdp]
\caption{The extracted fitting parameters from $\chi(T)$ data of $\mathrm{TlCuCl_3}$ ($T \geq 20$ K). $\chi_{0}$, $J$ and $\widetilde{J}$ were forced to be identical for the two field directions.  }
\begin{center}
\begin{ruledtabular}
\begin{tabular}{lclcl}
& $H \parallel b$ & $H \parallel [201]$ \\
\hline

$J/k_B$ (K)& \multicolumn{2}{c}{-61}&$\pm$1   \\
 $\widetilde{J}/k_B$ (K) & \multicolumn{2}{c}{-43.5}&$\pm$0.5  \\
 $\chi_{0}$ (m$^3$ /mol) & \multicolumn{2}{c}{-2.0 $\times$ 10$^{-9}$} &$\pm$10$^{-10}$ \\ 
  $g$ & 2.36  & 2.39&$\pm$0.05 \\ 

\end{tabular}
\end{ruledtabular}
\end{center}
\label{para}
\end{table}%

The value of the intradimer coupling $J$ is close to to the result obtained by neutron scattering measurements ($J/k_B \sim -64$ K \cite{cava}).
\\ Unfortunately, the fit does not allow us to distinguish between the individual interaction coupling constants contributing to $\widetilde{J}$ in $\mathrm{TlCuCl_3}$ as defined in Ref.\cite{cava}, but the fact that $\widetilde{J} \approx J$ clearly shows the strong 3D coupling between the dimers. From \eq{s} we find that the peak value $\chi_{max}=\chi(T \approx$ 36 K) increases with either increasing the $g$-factor or the intradimer coupling constant $J$, or by decreasing the interdimer coupling constant $\widetilde{J}$.  Since our value of $J$ is consistent with published data from neutron scattering measurements \cite{cava}, the slight overshoot of the fitting curve with respect to the measured data around $\chi_{max}$ implicates an underestimate of $\widetilde{J}$ and/or an overestimate of the Land\'e g-values, respectively. The latter scenario is supported by comparing our results to high precision ESR- measurements \cite{oosawa} which obtain a value of $g=$ 2.06 for both magnetic-field directions.

\section{\label{mag}The low-temperature magnetization}
 In the theory for a BEC of magnetic quasiparticles in insulating materials, the total magnetization (to be more precise, the total magnetic moment) $M=g \mu_B N$ is proportional to the total number of excited triplons $N$, which depends on both the temperature $T$ and magnetic field $H$  \cite{Nik}.  We therefore decided to analyze in detail the low-temperature region of the magnetization for both low magnetic fields (1 T $\leq \mu_0H \lesssim \mu_0H_c$) and high magnetic fields ($\mu_0H_c \lesssim \mu_0H \leq$ 9 T) using a consistent approach including adequate contributions for the respective magnetic field regions. We note here that all the magnetization data presented in this work are expressed as magnetic moment $M$ per single $\mathrm{Cu^{2+}}$ ion . The later used quantity $m(T)=M(T)/N_d=g\mu_Bn(T)$ (where $N_d$ is the number of dimers and $n(T)=N(T) / N_d$ is the total triplon density) differs from that by a factor 2. All values extracted from fits and calculations are presented in the latter units.
 
\subsection{\label{lowmag}The magnetization $M(T)$ for $\mathrm{1T} \leq \mu_0H \lesssim \mu_0H_c$}
\begin{figure}
\includegraphics[scale=0.3]{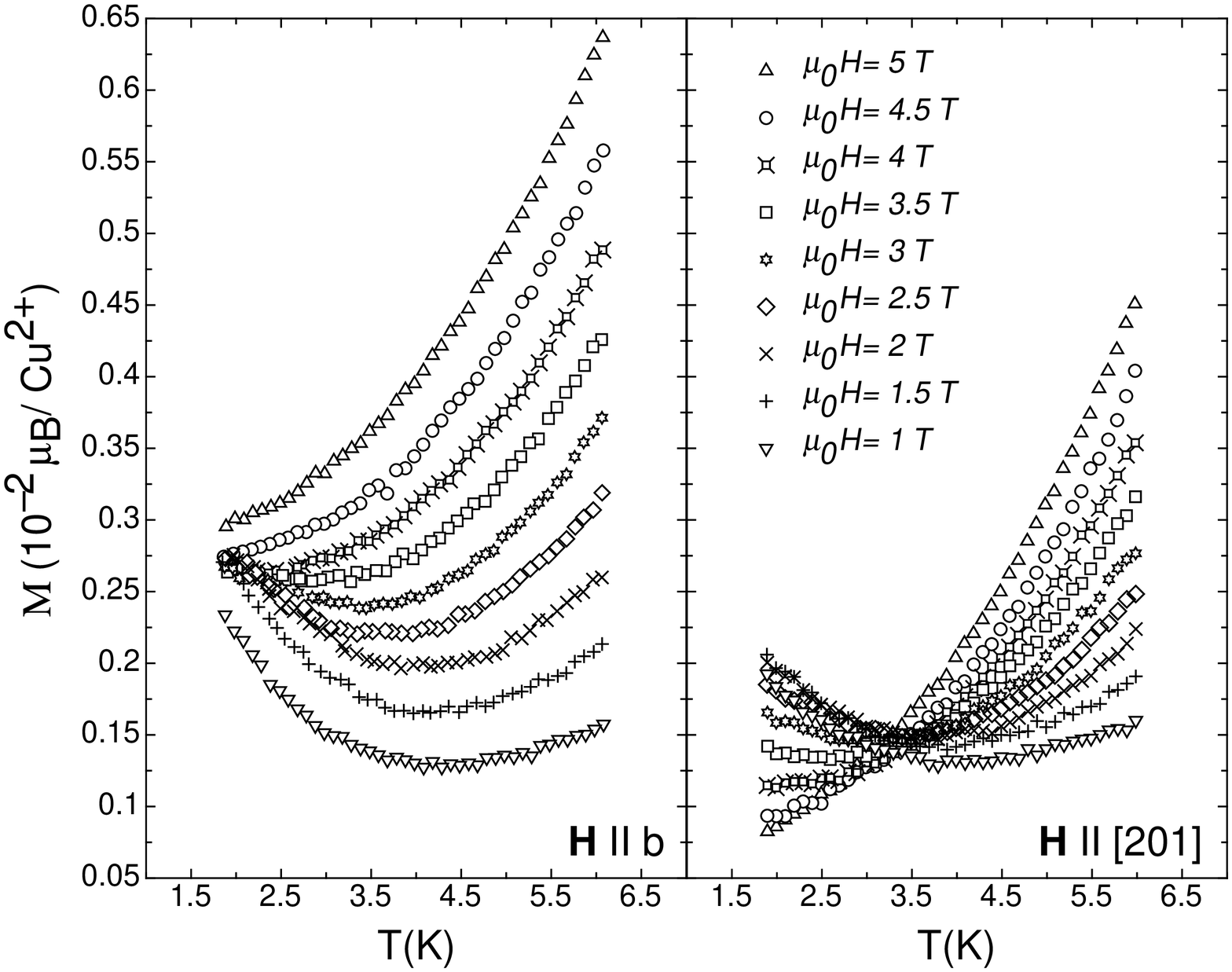}
\caption{\label{mlowboth} The magnetization $M(T)$ of $\mathrm{TlCuCl_3}$ for 1T $\leq \mu_0H \leq$ 5 T applied along the crystallographic $b$-axis (left) and along [201] (right).}
\end{figure}

\begin{figure}
\includegraphics[scale=0.3]{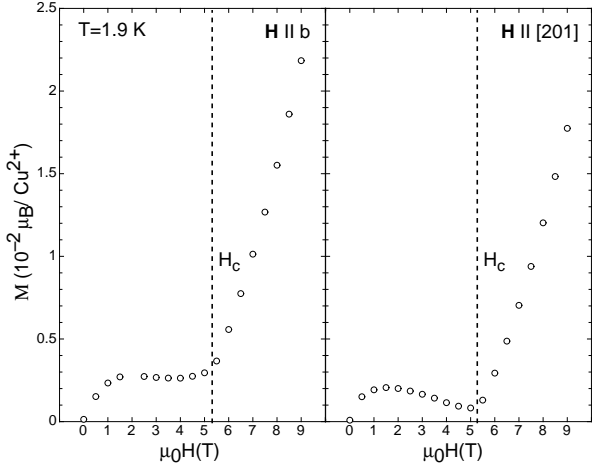}
\caption{\label{mhf} The magnetization $M(H)$ of $\mathrm{TlCuCl_3}$ for $T$ = 1.9 K with $H$ applied along the crystallographic $b$-axis (left) and along [201] (right).}
\end{figure}

Fig.~\ref{mlowboth} shows the variation of $M$ of $\mathrm{TlCuCl_3}$ at low temperatures along the crystallographic $b$- axis and the [201]- direction, respectively, for magnetic fields up to $\mu_0H=5$ T.  The magnetization decreases exponentially to almost zero with decreasing temperature for both crystallographic directions, but showing an upturn at low temperatures for low magnetic fields. With increasing magnetic field the anisotropic behavior of the magnetization in the two different field orientations becomes apparent. Because for both field directions the upturn in $M$ at low temperatures is gradually suppressed with increasing $H$,  the magnetization curves for $\mathbf{H} \parallel [201]$ cross at $T_{cross}\sim$ 3.2 K . For $\mathbf{H} \parallel b$ a similar crossing of $M(T)$ data cannot be seen in the analyzed temperature range, but by extrapolating the respective magnetization curves to lower temperatures a $T_{cross}$ below 2 K seems to be plausible. 
\\ This crossing of $M(T)$-data is caused by the fact that the upturn in $M(T)$ at low temperatures does not grow linearly with $H$. Moreover, this upturn is quantitatively different in the two considered field directions. If this low-temperature contribution was due to an extrinsic paramagnetic impurity phase it could be expected to be isotropic. We therefore consider this behavior to be intrinsic to $\mathrm{TlCuCl_3}$. This upturn in $M(T)$ can be expressed as a temperature and field-dependent Curie-Weiss-like term that is proportional to the Brillouin function $B_{S}\left(x\right)$ with $x=g\mu_B\mu_0HS/k_B T$ and a constant $C_{S}$. Because we assume this term to be intrinsic to the here studied dimer-system, it is reasonable to assign it to magnetic moments associated with the triplet states with $S=1$. A similar observation confirming this fact was reported in \cite{oosawa}. The magnetic-field dependence of our low-temperature data showing an almost saturated behavior in $M(H)$ for $H\to H_c$, see \fig{mhf},  can indeed be qualitatively well reproduced by a magnetization term that is proportional to a Brillouin function $B_{S}\left(x\right)$. The additional $H$-dependence as observed for $H<H_c$ and $\vec{H} \parallel [201]$ can be explained by taking again a diamagnetic term $m_{dia}=\chi_{0}\cdot H$ and an additional paramagnetic term $m_{HL}(H)$ (to be discussed below) into account. However, the quality of corresponding fits to our low-temperature $M(H)$-data does not allow us to clearly distinguish between $S$=1 and $S=\frac{1}{2}$.
Therefore, we will consider in the following both scenarios for the Curie-Weiss-like term $C_S\cdot B_S(x)$, and we will later argue that only the $S=1$ case fits to our data in a physically meaningful way.

From the expression for the free energy per unit length of a Heisenberg ladder \cite{troyer} 
\begin{subequations}
\begin{equation}
\label{free energy}
f=-\frac{k_B T}{2}\left[1+2 cosh\left(\frac{g\mu_B\mu_0HS}{k_B T}\right)\right] z(T)
\end{equation}
with
\begin{equation}
z(T)=\frac{1}{2\pi}\int_{-\pi}^{\pi}e^{-\frac{\varepsilon_{\mathbf{k}}}{k_BT}}dk,
\end{equation}
\end{subequations}
we can estimate the magnetization per dimer by multiplying $f$ from \eq{free energy} with a characteristic length $\bar{a}$. This quantity has been suggested to correspond to an average lattice constant $\bar{a}=(a\cdot b\cdot c\cdot sin \beta)^{(1/3)}=$ 0.79 nm \cite{yamada}, where $a$, $b$ and $c$ and $\beta=96.32 ^\circ$ are taken from crystallographic data of $\mathrm{TlCuCl_3}$ \cite{tanaka}.

\begin{figure}
\includegraphics[scale=0.3]{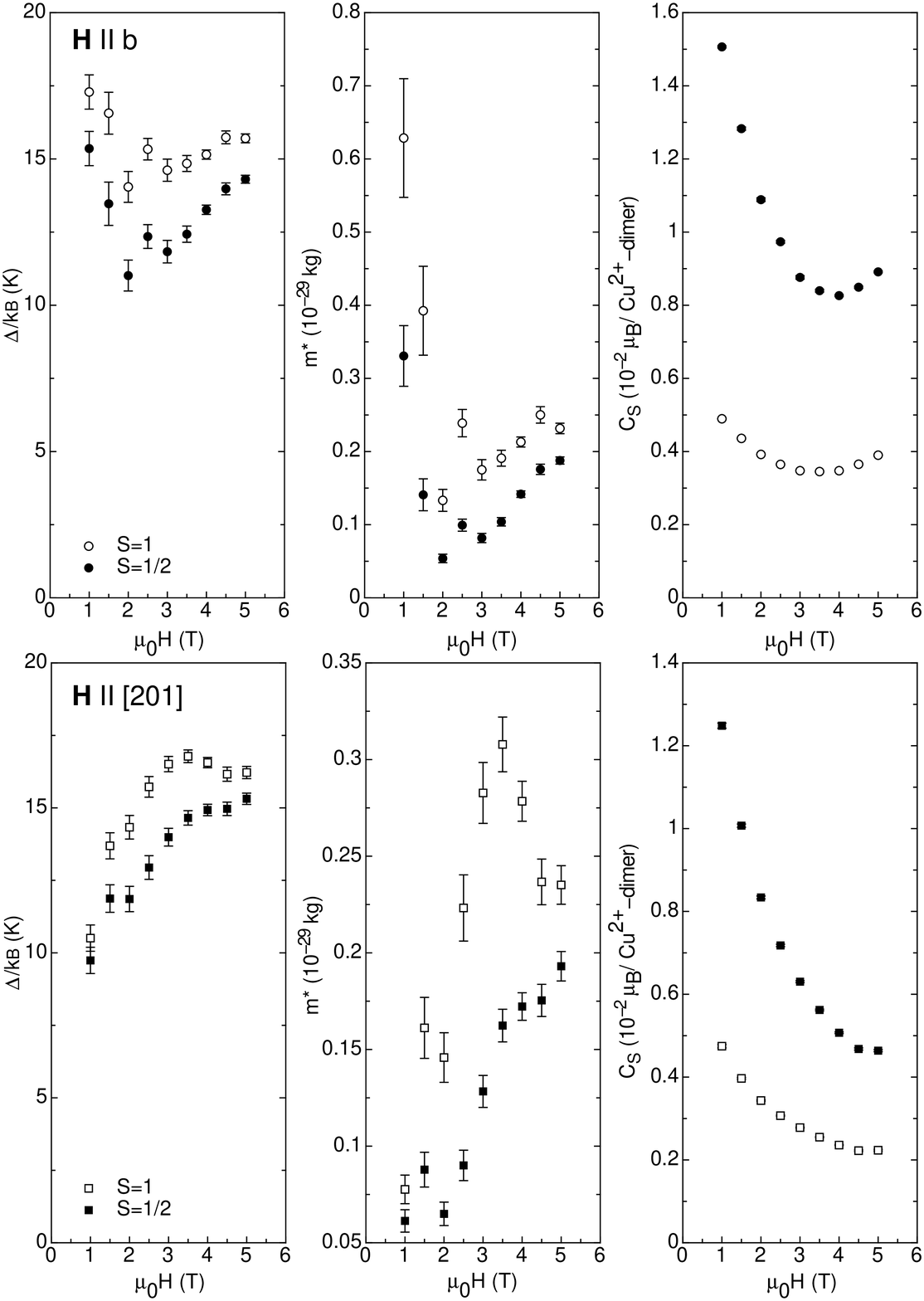}
\caption{\label{mlowall} The parameters $\Delta$, $m^{\star}$ and $C_S$ extracted from fits to the low-temperature magnetization data for $H \parallel b$-axis (upper frames) and for $H \parallel$ [201] (lower frames). Different scenarios for $m_{up}$ are indicated with open marks ($S$=1) and filled marks ($S=\frac{1}{2}$), respectively. The error bars of $C_S$ are of the order of the dot size.}
\end{figure}

Using a simple quadratic approximation for the triplon dispersion relation, $\varepsilon_{\mathbf{k}} \propto \Delta+\hbar^2k^2/2m^{\star}$ (where $m^{\star}$ corresponds to the effective mass of the triplons), one obtains \cite{troyer}
\begin{equation}
z(T)\approx\frac{1}{2\sqrt{\pi}} \left(\frac{\hbar^2}{2m^{\star}k_BT}\right)^{-\frac{1}{2}}e^{-\frac{\Delta}{k_B T}}.
\end{equation}
For the magnetization per dimer we therefore have
\begin{subequations}
\begin{equation}
\begin{aligned}
m_{HL}(T)&=-\bar{a} \frac{\partial f}{\partial H}\\
&= d \cdot \sqrt{T} e^{-\frac{\Delta}{k_B T}} sinh\left(\frac{g\mu_B\mu_0H}{k_B T}\right),\\
\end{aligned}
\end{equation}
with
\begin{equation}
\begin{aligned}
d &=g \mu_B \bar{a} \sqrt{\frac{k_Bm^{\star}}{2\pi \hbar^2}}.
\end{aligned}
\end{equation}
\end{subequations}
In order to analyze the upturn in $M(T)$ at low $T$, we include the above mentioned magnetization term 
\begin{equation}
m_{up}(T)= g \mu_BS\cdot C_{S} \cdot B_S\left(\frac{g\mu_B\mu_0H}{k_B T}S\right)
\end{equation}
for fixed magnetic field $H$.
We distinguish here between $S=\frac{1}{2}$ (non-intrinsic paramagnetic impurities) and the scenario $S=1$ (intrinsic term related to triplet states). We fitted the magnetization data at low enough temperatures ($T \leq$ 5 K) and 1 T$ \leq\mu_0H\leq$5 T for both field directions according to

\begin{equation}
\begin{aligned}
m(T)&=\frac{M(T)}{N_d}\\
&=m_{HL}(T)+m_{up}(T)+m_{dia}  \\
&=g\mu_B \bar{a} \sqrt{\frac{k_B m^\star}{2\pi \hbar^2}} \cdot \sqrt{T} e^{-\frac{\Delta}{k_B T}} sinh\left(\frac{g\mu_B\mu_0H}{k_B T}\right) \\
& \quad + g\mu_BS\cdot C_{S} \cdot B_S\left(\frac{g\mu_B\mu_0H}{k_B T}S\right) + m_{dia}
\end{aligned}
\end{equation}
with $g=2.06$.
\begin{figure}
\includegraphics[scale=0.3]{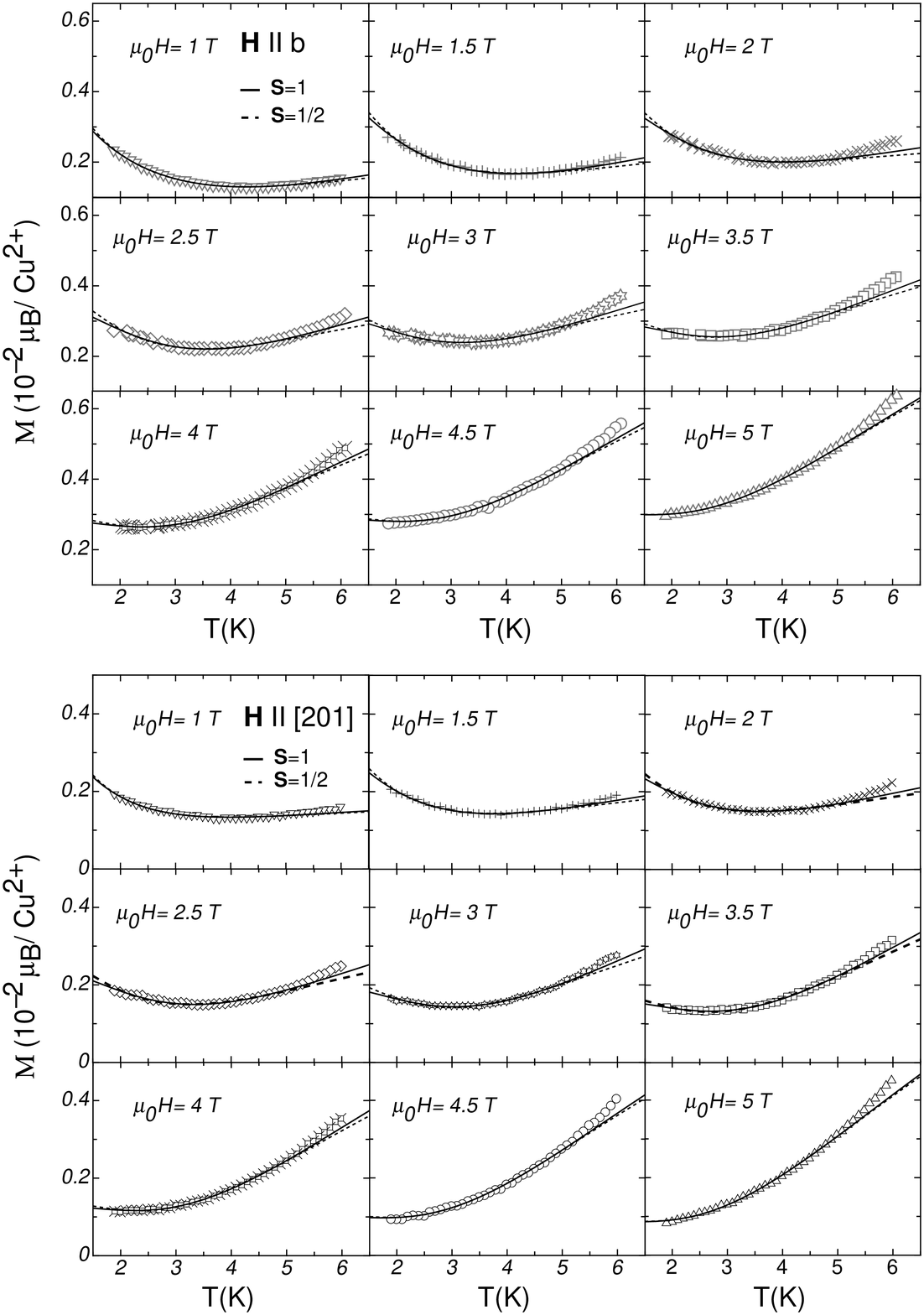}
\caption{\label{mlowallalongb} The magnetization $M(T)$ of $\mathrm{TlCuCl_3}$ for 1T $\leq \mu_0H \leq$ 5 T applied along the crystallographic $b$-axis (upper frames) and along [201] (lower frames) . The continuous and the dashed lines denote fits for $S=1$ and $S=\frac{1}{2}$, respectively (see text). }
\end{figure}
Because we were using the $T$-independent diamagnetic contribution $m_{dia}=\chi_0\cdot H$ extracted from the high-temperature susceptibility fits presented above, only the gap $\Delta$, the constant $C_{S}$ (for $S$=1 or $\frac{1}{2}$) and  the effective mass of a triplon $m^{\star}$ were fitting parameters. The corresponding fits to the magnetization data are shown in \fig{mlowallalongb}, while the corresponding results for the fitting parameters are presented in \fig{mlowall}. 
The values for the gap $\Delta$ slightly vary with magnetic field for both field directions around $\Delta/k_B\approx 13$ K, which is somewhat larger than $\Delta\approx 0.7$ meV = 8.3 K \cite{cava} determined by neutron scattering.  The triplon mass $m^\star\approx 0.2 \cdot 10^{-29}$ kg is an order of magnitude smaller than compared to the results from calculations and a corresponding analysis of high-~field magnetization data within the Hartree-Fock approximation \cite{yamada}. This discrepancy might be explained by our choice of $\bar{a}$ or by the use of the simplified quadratic energy-dispersion relation for this temperature region. The range of validity of a quadratic approximation is indeed restricted to lower temperatures ($T<$1 K) \cite{troyer,misguich, sirker} that are not accessible in our experiment.

The Curie-like contribution $C_{S}$ decreases for both cases $S$=1 and $S=\frac{1}{2}$ with increasing magnetic field $H$. For both field directions $C_{S}(H)$ shows a similar trend, although the variation with $H$ is much less pronounced for $S$=1. 
Since $C_S$ is expected to be a constant for a given magnetic-field direction, this fact already here strongly supports an $S=1$ scenario for a correct description of the paramagnetic background. We may speculate that this Curie-like term with $S$= 1 comes from a contribution of defects in the crystal or from dimers that are situated near the crystal boundaries.

\subsection{\label{highmag}The magnetization $M(T)$ for $\mathrm{5.5 T \leq \mu_0H \leq 9 T}$}
\begin{figure}
\includegraphics[scale=0.3]{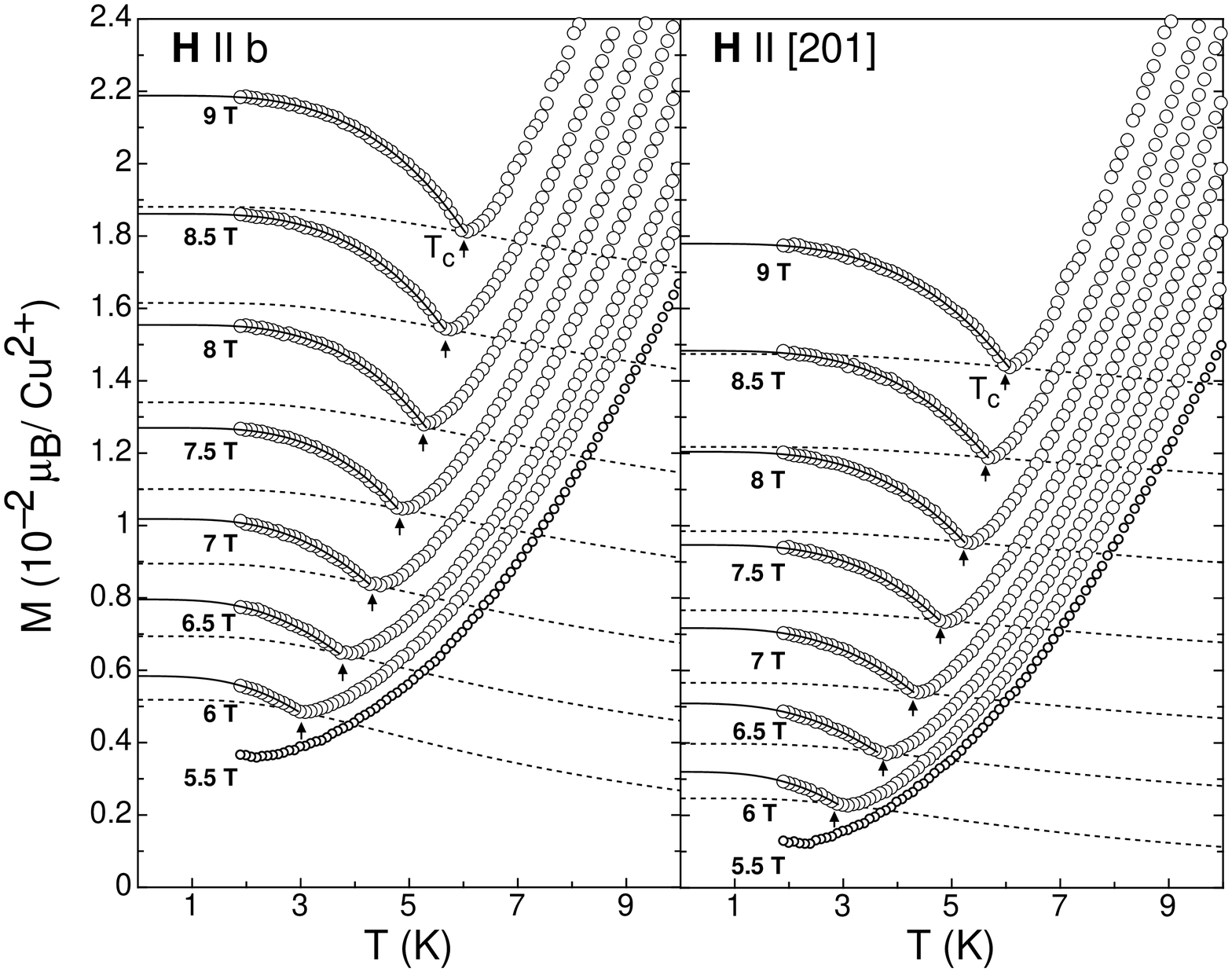}
\caption{\label{mhighallb} The magnetization $M(T)$ of $\mathrm{TlCuCl_3}$ for 5.5 T $\leq \mu_0H \leq$ 9 T applied along the $b$-axis (left) and along [201] (right), respectively. The critical temperature $T_{c}$ is marked by arrows. The solid lines correspond to fits to the data according to \eq{fappr} with $S=1$ for ($T<T_c$). The dashed lines represent calculated $M(T)$-curves using \eq{fappr} and the  same fitting parameters but with $n_0$=0.}
\end{figure}

\begin{figure}
\includegraphics[scale=0.3]{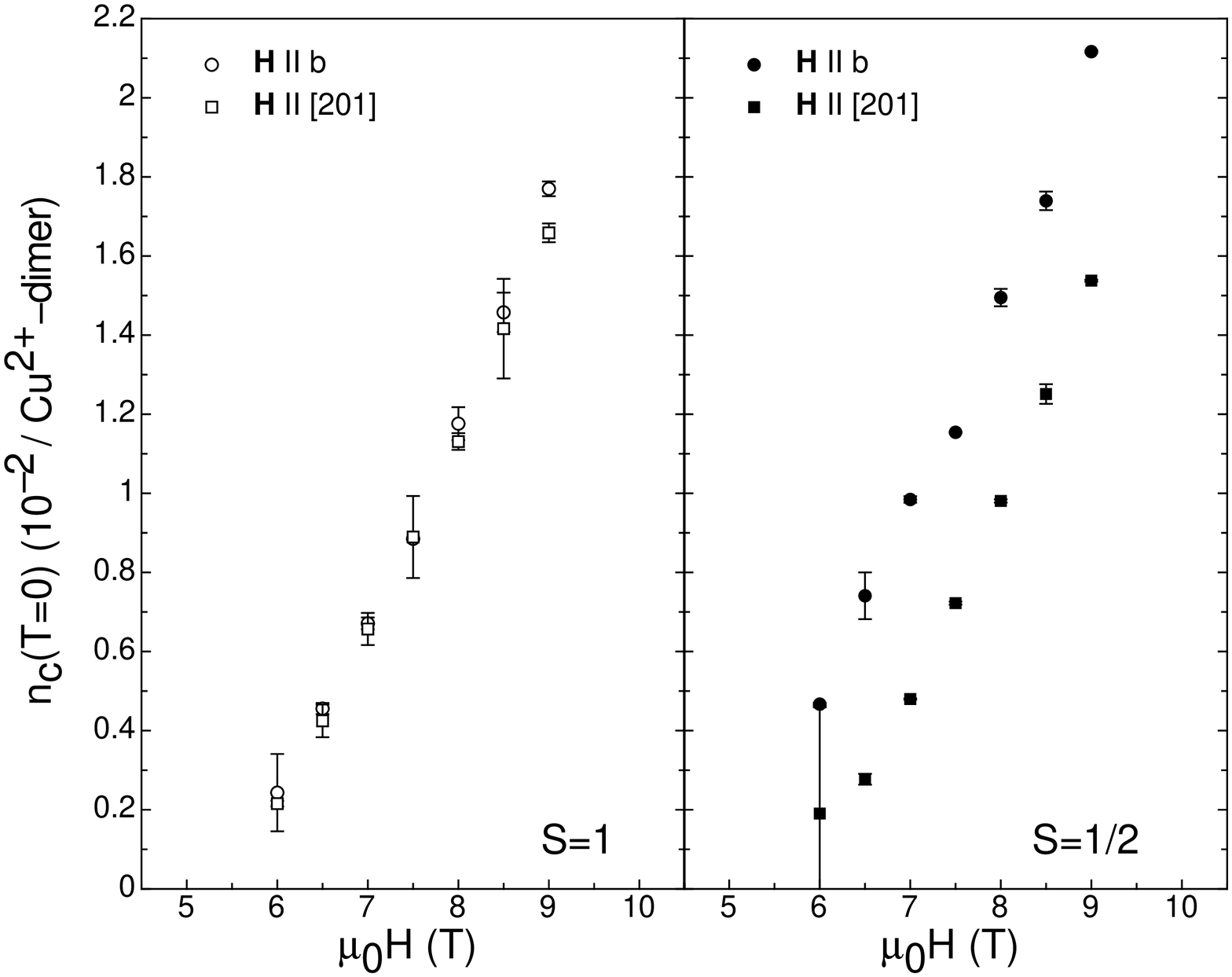}
\caption{\label{n} The condensate density $n_c(T=0)$ with different scenarios for $m_{up}$ (left: $S=1$, right: $S=\frac{1}{2}$).}
\end{figure}

\begin{figure}
\includegraphics[scale=0.3]{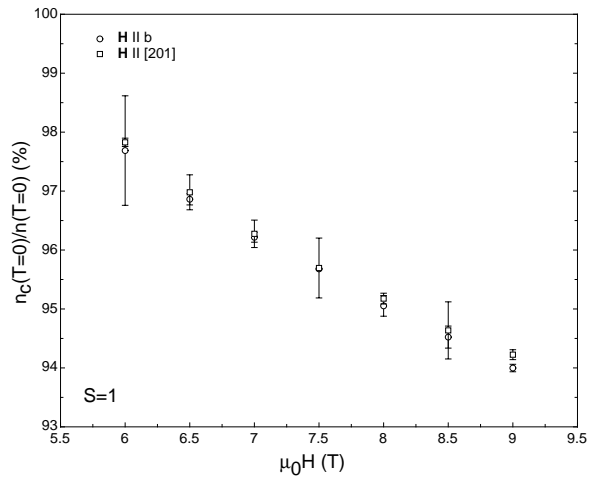}
\caption{\label{alln} The triplon fraction $n_c(0)/n(0)$ forming the condensate in the $S=1$ scenario for $m_{up}$.}
\end{figure}

The temperature dependence of the magnetization $M(T)$ along the applied magnetic field $\mathbf{H}$ shows a cusp-like minimum at a critical temperature $T_c(H)$ for fixed magnetic fields $H \geq H_c$, see \fig{mhighallb}. The increase of $M$ for $T<T_c$ is a consequence of the condensation of the magnetic quasiparticles and the increasing number of particles $N_c$ in the ground state forming the condensate. Theoretical arguments suggest within a simplified model a $T$-dependence of  $M \propto \left(1-\frac{T}{T_c}\right)^\frac{3}{2}$ for $T<T_c$ \cite{Nik} which is not observed in the experimental data, however.
\\ At high magnetic fields we have therefore fitted the low-temperature magnetization per dimer $m(T)$ according to a more general power-law including the diamagnetic contribution that we extracted from high-temperature magnetic-susceptibility measurements, and again a net paramagnetic moment $m_{up}(T,H)$ assumed to be proportional to the Brillouin function $B_{S}(x)$ for $S=1$ and $S=\frac{1}{2}$, respectively. For 6 T$\leq\mu_0H\leq$9 T we use
\begin{equation}
\label{fappr}
\begin{aligned}
m(T)&=\frac{M(T)}{N_d}\\
&=g \mu_B\frac{N(T)}{N_d}+m_{up}+m_{dia}   \\
&= g \mu_B (n_{crit}+n_0 \left( 1-\left(\frac{T}{T_c}\right)^\alpha \right))\\
& \quad + g\mu_BS \cdot C_{S}(H) \cdot B_S\left(\frac{g\mu_B\mu_0H}{k_B T}S\right) + m_{dia}
\end{aligned}
\end{equation}
where $n_{crit}=~N(T=T_c)/N_d$ is the critical density at which condensation occurs, corresponding to the normalized magnetization $m(T=T_c)=g \mu_B n_{crit}$. The physical meaning of the exponent $\alpha$, see \fig{allpar}(b), is not further discussed here, see Appendix.
\\ At zero temperature we have for fixed magnetic field $H$
\begin{equation}
\begin{aligned}
m(T=0)&=g \mu_B(n_{crit}+n_0)+m_{up}(T=0)+m_{dia}\\
&=g \mu_Bn(0)+m_{up}(T=0)+m_{dia}
\end{aligned}
\end{equation} 
with the total triplon density at $T=0$,  $n(0)=n_{crit}+n_0$ .
\\ For an ideal Bose gas $n(0)$ corresponds to the condensate density $n_c(0)$. As soon as interactions between the particles are considered, the depletion of the condensate has to be taken into account. The quantity 
\begin{equation}
\label{sum}
n(0)=n_c(0) +\tilde{n}(0)
\end{equation} 
is then a sum of the condensate density $n_c(0)$ and the density of noncondensed particles $\tilde{n}(0)$. The latter term represents the number of triplons per Cu$^{2+}$ dimer scattered out of the ground state due to the interactions between the particles. It depends on the number of condensed particles and can be expressed as \cite{Nik}
 \begin{equation}
 \label{ex}
 \tilde{n}(0)= \frac{1}{3\pi^2}\left( \frac{m^{\star} U_0 n_c(0)}{\hbar^2}\right)^\frac{3}{2}, 
 \end{equation} 
 where we take the mass of a triplon $m^\star \approx 2.6 \times 10^{-29}$ kg and the two-particle interaction potential $U_0/k_B\approx$ 315 K from Ref.\cite{yamada}. 
 Replacing $\tilde{n}(0)$ in \eq{sum} with the expression in \eq{ex} we obtain
 \begin{equation}
 \label{sumsum}
 n(0)=n_c(0) +\frac{1}{3\pi^2}\left( \frac{m^{\star} U_0 n_c(0)}{\hbar^2}\right)^\frac{3}{2}.
 \end{equation}  
From our fits according to \eq{fappr} and with $n(0)=n_{crit}+n_0$ we can now calculate the condensate density at zero temperature $n_c(0)$ for various magnetic fields using \eq{sumsum}, see \fig{allpar}(a).
As one would expect from simple arguments \cite{Nik, yamada} $n_c(0)$ increases with increasing magnetic field. 
It is essential to note that the number of triplons $N_c(0)=n_c(0)\cdot N_d$ forming the condensate at $T=0$ is the same for both field directions only in the $S=1$ scenario for $m_{up}(T)$, see \fig{n}, and only in this scenario $n_c(0)$ extrapolates to zero at the correct critical field $\mu_0H_c \approx$ 5.5 T. These facts again strongly support our hypothesis that $m_{up}(T)$ is intrinsic with $S=1$, and it confirms the interpretation of the magnetic field $\mathbf{H}$ acting as the chemical potential \cite{giaandrueg}.
\\ In $\mu_0H=$ 6 T, right above the critical field $H_c$, the percentage of the condensed particles $n_c(0)$ with respect to the total density of triplons $n(0)$ is approximately 98\% and slightly decreases with increasing magnetic field, see \fig{alln}. This result is consistent with a similarly low noncondensed magnon density as calculated in Ref. \cite{sirker}, where $\tilde{n}(0)$ increases from zero for $H = H_c$ to approximately 7\% of the total triplon density at $T$ = 0 in $\mu_0H =$ 7 T. From the high percentage of condensed particles we can confirm that the triplons in $\mathrm{TlCuCl_3}$ form a \textit{weakly interacting Bose gas} \cite{sirker} right above $H_c$, and that the interaction increases with increasing particle density, i.e. with increasing the magnetic field $H$.
\begin{figure}
\includegraphics[scale=0.3]{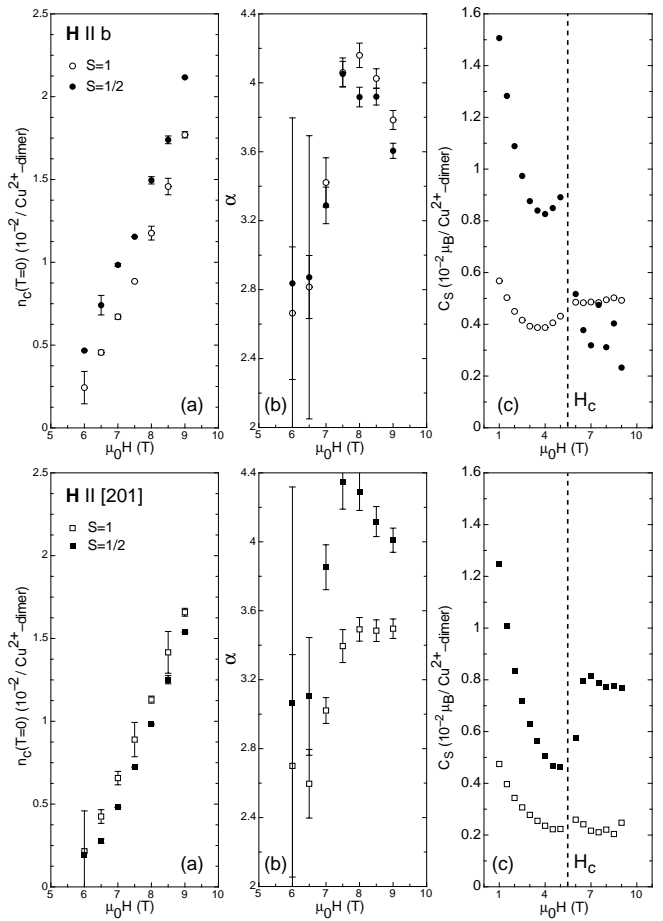}
\caption{\label{allpar} The condensate density $n_c(T=0)$ (a), the exponent $\alpha$ (b) and the Curie-~like contribution $C_S$ (c) for $H \parallel b$ (upper frames) and for $H \parallel$ [201] (lower frames). Different scenarios for $m_{up}$ are indicated with open marks ($S$=1) and filled marks ($S=\frac{1}{2}$), respectively. The error bars of $C_S$ are of the order of the dot size.}
\end{figure}
\\Finally, we want to mention that the Curie-like contribution $C_S$ is small and essentially constant for $H > H_c$ in both the $S=\frac{1}{2}$ and the $S=1$ scenarios, see \fig{allpar}(c). However, the corresponding data for $S=1$ are more or less smooth continuations of the respective data for $H<H_c$, in very contrast to the $C_S$ data for $S=\frac{1}{2}$ that show a discontinuity around $H=H_c$, see \fig{allpar}(c). The comparably moderate variation of $C_S$ with $H$ over the whole considered range of magnetic fields for the $S=1$ scenario (covering both the normal phase and the BEC obeying an entirely different physics) may indicate that $C_S$ is indeed a constant for each magnetic-field direction, and that our interpretation of a $S=1$ paramagnetic background is correct.

\section{Conclusions}
We have presented an analysis of magnetization $M(T,H)$-data of $\mathrm{TlCuCl_3}$ and we calculated the density of condensed particles $n_c(0)$ at $T=0$. The percentage of $n_c(0)$ with respect to the total density of triplons $n(0)$ is approximately 98\% in $\mu_0H=$ 6 T and slightly decreases with increasing magnetic field. We demonstrated that this fraction is the same for both $\vec{H} \parallel b$ and $\vec{H} \parallel [201]$ if we assume the presence of a small number of intrinsic $S=1$ magnetic moments that are not part of the Bose-Einstein condensate of triplons even at the lowest temperatures.

\begin{acknowledgments}
This work was supported by the Schweizerische Nationalfonds zur F\"orderung der wissenschaftlichen Forschung, Grant. No. 20-111653.
\end{acknowledgments}

\begin{figure}
\includegraphics[scale=0.3]{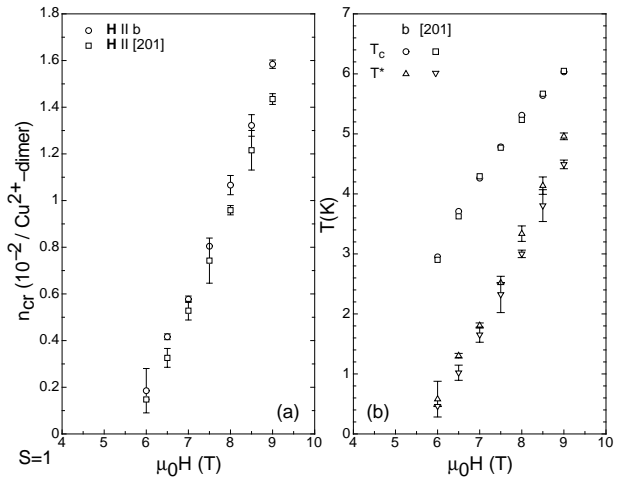}
\caption{\label{temp} The critical triplon density $n_{cr}$ (a) and the corresponding typical temperatures $T^{\ast} \approx n_{cr} U_0/ k_\mathrm{B}< T_c$ (b) in the $S=1$ scenario for $m_{up}$ as functions of the applied magnetic field for $\vec{H} \parallel b$ (circles and upward triangles) and $\vec{H} \parallel [201]$ (squares and downward triangles), respectivley.}
\end{figure}


\section{Appendix}

We want to emphasize that we do not interpret the exponent $\alpha$ in the power-law approach (\eq{fappr}) used for fitting the low-temperature magnetization data at high magnetic fields as a universal critical exponent. In this sense its physical meaning is not clear. 
\\ In general, the normalfluid density in a dilute Bose gas in the condensed phase is proportional to $T^4$ at low enough temperatures  ($T \ll {T^{\ast}} \approx n U_0/k_\mathrm{B}$), where $n$ is the total particle density  and $U_0$ the interaction energy.
\\ In the case of $\mathrm{TlCuCl_3}$ we can estimate $T^{\ast}$ by replacing $n\cong n_{cr}$ and $U_0/k_B\approx$~315~K \cite{yamada}, see \fig{temp}. 
The values for $\alpha$ extracted from the fits, see \fig{allpar}(b), vary around $\alpha \sim$ 4 with a decreasing tendency and increasing fitting error as $H \to H_c$. 
This can be explained by the fact that for high magnetic fields $T^{\ast}$ is fairly close to $T_c$, whereas for low magnetic fields the difference between the two characteristic temperatures increases, thereby restricting the validity of the $T^4$-~power law to very low temperatures that are not accessible in our experiment.
Nevertheless, it is clear that the increasing magnetization for $T\to0$ is related to the increasing fraction of the condensed particles $n_c(T)$. The evaluation of $n(0)$ from our phenomenological power law, see \eq{fappr}, and the calculation of $n_c(0)$ using \eq{sumsum} gives, in any case, a reliable estimate of the intercept of $n_c(T)$ at $T=0$, irrespective of the correct functional form of $n_c(T)$.






\bibliography{dellamorebiblio}

\end{document}